\documentclass[12pt]{article}
\usepackage{amsmath,amsfonts,amssymb,amsthm}
\usepackage{pdfpages}
\usepackage{graphicx}
\usepackage{subfigure}
\newtheorem{proposition}{Proposition}[section]
\newtheorem{theorem}[proposition]{Theorem}

\newcommand{\nc}{\newcommand}
\nc{\I}{{\mathbf 1}}

\nc{\R}{{\mathbb R}}
\nc{\N}{{\mathbb N}}
\nc{\Z}{{\mathbb Z}}

\nc{\PP}{\mathbb{P}}
\nc{\BE}{\mathbb{E}}
\nc{\BV}{\mathbb{V}}
\numberwithin{equation}{section}

\begin{document}

\renewcommand{\thefootnote}{\fnsymbol{footnote}}
\author{{\sc Norbert Henze\footnotemark[1]\, and Celeste Mayer\footnotemark[2]}}
\footnotetext[1]{norbert.henze@kit.edu,
Karlsruhe Institute of Technology, Institute of Stochastics,
76131 Karlsruhe, Germany. }
\footnotetext[2]{celeste.mayer@kit.edu,
Karlsruhe Institute of Technology, Institute of Stochastics,
76131 Karlsruhe, Germany. }

\title{More good news on the (only) affine invariant test for multivariate reflected symmetry about an unknown center}
\date{\today}
\maketitle

\abstract{We revisit the problem of testing for multivariate reflected symmetry about an unspecified point. Although this testing
problem is invariant with respect to full-rank affine transformations, among the hitherto few proposed tests only the test studied in \cite{4}
 respects this property. We identify a measure of deviation $\Delta$ (say) from symmetry associated with the test statistic $T_n$ (say),
 and we obtain the limit normal distribution of $T_n$  as $n \to \infty$ under a fixed alternative to symmetry. Since a consistent estimator of the variance of this limit
normal distribution is available, we obtain an asymptotic confidence interval for $\Delta$. The test, when applied to a classical data set,
 strongly rejects the hypothesis of reflected symmetry, although other tests even do not object against the much stronger hypothesis of elliptical symmetry.
  }



\section{Introduction}

Testing for symmetry of a univariate distribution about a specified or unspecified point has been a topic of intensive research,
see e.g., Section 3 of \cite{quessy}.  In the multivariate case, this problem is more complex, since  different notions of symmetry are available.
Among these are, in increasing order of specialization, {\em reflected} ({\em diagonal}) {\em symmetry}, {\em spherical symmetry}, and {\em elliptical symmetry}, see, e.g.,
  \cite{meintanis12} for an account on the importance of the assumption of symmetry and
  a survey on these concepts and corresponding goodness-of-fit tests.

  In this paper, we consider testing for reflected symmetry. To be specific, let $X,X_1,X_2,\ldots $ be a sequence of independent and identically distributed (i.i.d.) $d$-dimensional random (column) vectors, defined on some common probability space $(\Omega,{\cal A},\PP)$, and assume $d \ge 1$.  Thus, the univariate case  is deliberately not excluded in what follows. Writing $\overset{\mathcal{D}}{=}$ for equality in distribution, the problem is to test the hypothesis
   \begin{equation}\label{testproblem}
H_0: X- \mu \ \overset{\mathcal{D}}{=} \ \mu-X \text{ for some (unknown) } \mu \in \R^d,
\end{equation}
of {\em reflected} ({\em diagonal}) {\em symmetry} about an unspecified point, against general alternatives.

The technically less demanding problem of testing for reflected symmetry about a {\em specified} point has been considered in \cite{aki} and, in the special case $d=2$,
in \cite{dyckerhoff} and \cite{einmahlgan}. For distributions concentrated on the unit circle, the hypothesis "$X \overset{\mathcal{D}}{=} -X$" is called
{\em circular reflective symmetry}, see \cite{leyetal} and the references therein. Symmetry of a bivariate distribution about a given line is studied in
\cite{raorag}.

Notice that if a test of $H_0$ rejects the hypothesis of reflected symmetry, it is forced to also reject the stronger hypotheses of spherical or elliptical symmetry. Thus, any test of $H_0$ is in this sense a  "necessary test" for spherical or elliptical symmetry, and even for multivariate normality.

There is a further basic issue inherent in the testing problem (\ref{testproblem}). Suppose $X- \mu \overset{\mathcal{D}}{=} \mu-X$, and let
$A$ be a regular ($d\times d$)-matrix and $b \in \R^d$. Then
\[
AX+b - (A\mu +b) \overset{\mathcal{D}}{=} A\mu + b - (AX+b).
\]
This means that the problem of testing for reflected symmetry about an unspecified point
is invariant with respect to full rank affine transformations of $X$.
As a consequence, any genuine test of $H_0$ based on $X_1,\ldots,X_n$ should respect this property.
Hence, if $T_n = T_n(X_1,\ldots,X_n)$ is a test statistic based on $X_1,\ldots,X_n$, we should have affine invariance of $T_n$, i.e.,
\[
T_n(AX_1+b, \ldots,AX_n +b) = T_n(X_1,\ldots,X_n)
\]
for each nonsingular $A \in \R^{d\times d}$ and $b \in \R^d$. Although there are a few attempts to tackle the problem (\ref{testproblem})
(see \cite{szekely}, \cite{Heathcote}, \cite{Neuhaus},  \cite{nga} and Section 2.1 of \cite{meintanis12}), none of the proposed test statistics is affine
invariant. It is the purpose of this paper to revisit the test of Henze, Klar and Meintanis \cite{4}. This test
is affine invariant for the testing problem (\ref{testproblem}), easy to use, consistent against general alternatives, and it is able to detect alternatives that approach the hypothesis at the rate $n^{-1/2}$. We sum up these (and more) properties in Section \ref{sechkm}. In Section
 \ref{alter}, we consider a fixed alternative distribution to $H_0$  and identify a measure of deviation $\Delta$ (say) from symmetry associated with
  the test statistic of \cite{4}. Moreover, we prove that the test statistic has a limit normal distribution.
  In Section \ref{secesti}, we present a consistent estimator of the variance of this limit distribution, which yields an asymptotic confidence interval for
  $\Delta$. Section \ref{secexam} presents two examples, whereas Section \ref{secrealdata} applies the  test to a data set from a health survey of paint sprayers in a car assembly plant.
For the sake of readability, most of the proofs are deferred to Section \ref{secproofs}.

\section{The HKM-test}\label{sechkm}
The test of Henze et al. \cite{4} -- henceforth termed the HKM-test -- rejects $H_0$ for large values of the test statistic
  \begin{equation*}
T_{n,a}  =  \int_{\R^d}\left(\frac{1}{\sqrt{n}}\sum_{j=1}^{n}\sin\left(t^\top  Y_{n,j}\right)\right)^2\exp\left(-a\|t\|^2\right)\textrm{d}t,
\end{equation*}
where $a>0$ is some fixed parameter. Here, $\top$ denotes transposition of vectors and matrices, $\| \cdot \|$ is the Euclidean norm in $\R^d$,
\begin{equation}\label{defynj}
Y_{n,j} = S_n^{-1/2}(X_j - \overline{X}_n), \quad j=1,\ldots,n,
\end{equation}
are the {\em scaled residuals} of $X_1,\ldots,X_n$, and $\overline{X}_n = n^{-1} \! \sum_{j=1}^n \! X_j$, $S_n = n^{-1}
\! \sum_{j=1}^n (X_j-\overline{X}_n)(X_j - \overline{X}_n)^\top $ denote the sample mean and the sample covariance matrix of $X_1,\ldots,X_n$,
respectively. The matrix $S_n^{-1/2}$ is the unique symmetric square root of $S_n^{-1}$.  To ensure the almost sure invertibility of $S_n$, we make the basic tacit assumptions that the distribution of $X$ (henceforth abbreviated by $\PP^X$) is absolutely continuous with respect to Lebesgue measure, and that $n \ge d+1$, see \cite{12}.

An alternative representation of
$T_{n,a}$ is
\begin{equation*}
T_{n,a}  =   \frac{\pi^{d/2}}{2na^{d/2}}\sum_{i,j=1}^{n}\left[\exp\left(-\frac{1}{4a}\|Y_{n,i}-Y_{n,j}\|^2\right)-\exp\left(-\frac{1}{4a}\|Y_{n,i}+Y_{n,j}\|^2\right)\right],
\end{equation*}
which is amenable to computational purposes. Notice that $T_{n,a}$ is a function of the Mahalanobis angles and distances
$Y_{n,i}^\top  Y_{n,j} = (X_i- \overline{X}_n)^\top  S_n^{-1} (X_j- \overline{X}_n)$, $i,j=1,\ldots,n$, and is thus
affine invariant (see also Section 2 of \cite{henze2002}). Besides, it is not necessary to compute the square root of $S_n^{-1}$.

A further representation of $T_{n,a}$ is
\begin{equation}\label{tna3}
T_{n,a} = \frac{n(2\pi)^d}{4}\int_{\R^d}\left(\widehat{f}_{n,a}(x)-\widehat{f}_{n,a}(-x)\right)^2\textrm{d}x,
\end{equation}
where
\begin{equation}\label{kerneld}
 \widehat{f}_{n,a}(x)=\frac{1}{n}\sum_{j=1}^{n}\frac{1}{(2\pi a)^{d/2}}\exp\left(-\frac{\|x-Y_{n,j}\|^2}{2a}\right)
\end{equation}
is a nonparametric kernel density estimator of $f$ with Gaussian kernel $(2 \pi)^{-d/2}\textrm{e}^{-\|x\|^2/2}$ and bandwidth $a$.

Some more light on the role of $a$ is cast by the relation
\begin{equation}\label{limitskew}
\lim_{a\rightarrow \infty}\frac{96}{n\pi^{d/2}}a^{d/2+3}T_{n,a}=2b_{n,1}+3b_{n,2}.
\end{equation}
Here, the limit is elementwise on the underlying probability space, and
\[ b_{n,1}=\frac{1}{n^2}\sum_{i,j=1}^{n}\left(Y_{n,i}^\top  Y_{n,j}\right)^3, \quad
b_{n,2}=\frac{1}{n^2}\sum_{i,j=1}^{n}Y_{n,i}^\top  Y_{n,j}\left\|Y_{n,i}\right\|^2\left\|Y_{n,j}\right\|^2
\]
denote empirical multivariate skewness in the sense of Mardia \cite{Mardia} and M\'{o}ri et al. \cite{MoriMori}, respectively.
Thus, for large values of $a$, the test statistic $T_{n,a}$, apart from a scaling factor,  is approximately a linear combination of two
measures of skewness. In the univariate case $b_{n,1}$ and $b_{n,2}$ coincide, and \eqref{limitskew} specializes to give
\[
\lim_{a\rightarrow \infty}\frac{96}{n\sqrt{\pi}}a^{7/2}T_{n,a}=5 \left(\frac{1}{n} \sum_{j=1}^n \left(\frac{X_j - \overline{X}_n}{s_n} \right)^3 \right)^2,
\]
where $s_n^2 = n^{-1}\sum_{i=1}^n (X_i- \overline{X}_n)^2$.

If $\BE \|X\|^4 < \infty$, we have
\begin{equation*}
 T_{n,a} \overset{\mathcal{D}}{\longrightarrow}\int_{\R^d}\mathcal{W}^2(t)\exp\left(-a\|t\|^2\right)\textrm{d}t,
  \end{equation*}
  under $H_0$, where $\overset{\mathcal{D}}{\longrightarrow}$ denotes convergence in distribution, and
    $\mathcal{W}$ is some centred Gaussian process in the Hilbert space $\mathcal{L}^2 =\mathcal{L}^2\left(\R^d,\mathcal{B}^d,  \exp\left(-a\left\|t\right\|^2\right)\right)$ of (equivalence classes of) measurable functions $f:\R^d \to \R$ that are
square integrable with respect to the measure $\exp\left(-a\left\|t\right\|^2\right)  \textrm{d}t$.

Under a triangular array $X_{n,1},\ldots,X_{n,n}, n \ge d+1$,  of row-wise
i.i.d. random vectors with density
\[
f_n(x) = f_0(x) \left( 1+ \frac{h(x)}{\sqrt{n}}\right), \quad x \in \R^d,
\]
where $f_0$ is a density symmetric about $0$, and $h$ is a bounded measurable function satisfying $\int_{\R^d} h(x)f_0(x) \textrm{d} x =0$, we have
\[
 T_{n,a} \overset{\mathcal{D}}{\longrightarrow}\int_{\R^d}\left(\mathcal{W}(t)+s(t)\right)^2 \exp\left(-a\|t\|^2\right) \textrm{d}t,
 \]
 where
 \[ s(t)  =  \int_{\R^d}\left[\sin\left(t^\top  x\right)-t^\top  \psi(t)x\right]h(x)f_0(x)\, \textrm{d}x, \quad
 \psi(t) =  \int_{\R^d} \cos(t^\top  x) f_0(x) \, \textrm{d} x.
\]
Hence, the test has positive asymptotic power against contiguous alternatives that approach the null hypothesis at the rate $n^{-1/2}$.

Since both the finite-sample and the limit null distribution of $T_{n,a}$ depend on the unknown distribution of $X$, the test is carried out
as permutation test. To this end, let $U_1,U_2, \ldots $ be a sequence of i.i.d. random variables, independent of $X_1,X_2,\ldots$, such that
$\PP(U_j=1) = \PP(U_j=-1)= 1/2$. Conditionally on $Y_{n,j} =y_j$, $j=1,\ldots,n$, let $Z_j = U_jy_j$, $j=1,\ldots,n$ and put
$\overline{Z}_n = n^{-1}\sum_{j=1}^n Z_j$.  \cite{4} shows that the permutation statistic
\[
T_{n,a}^P = \int_{\R^d} \left({\cal W}_n^P(t)\right)^2 \, \exp(-a \|t\|^2) \, \textrm{d}t,
\]
which is based on the so-called {\em permutation process}
\[
{\cal W}_n^P(t) = \frac{1}{\sqrt{n}} \sum_{j=1}^n U_j \left\{ \sin(t^\top  y_j) - \left( \frac{1}{n} \sum_{k=1}^n \cos(t^\top  y_k)\right) t^\top  y_j \right\},
\]
takes the form
\begin{eqnarray*}
T_{n,a}^P & = & \frac{\pi^{d/2}}{2a^{d/2}n} \sum_{i,j=1}^n \left[ \left( 2 + \frac{\|\overline{Z}_n\|^2}{2a} - \left\{ 1+
\frac{(Z_i-Z_j)^\top \overline{Z}_n}{2a} \right\}^2 \right) \exp \left( - \frac{\|Z_i-Z_j\|^2}{4a}\right) \right. \\
&  & \ \quad + \left. \left( \frac{\|\overline{Z}_n\|^2}{2a} - \left\{ 1+ \frac{(Z_i+Z_j)^\top \overline{Z}_n}{2a} \right\}^2 \right)
\exp\left( - \frac{\|Z_i+Z_j\|^2}{4a}\right) \right].
\end{eqnarray*}
Moreover, the limit distribution of $T_{n,a}^P$ under $H_0$ is the same as that of $T_{n,a}$
for almost all sample sequences $X_1,X_2, \ldots$ .
Under a fixed alternative distribution satisfying $\BE \|X\|^2 < \infty$ (which, in view of affine invariance, is assumed to have zero expectation and unit covariance matrix), we have
\[
\lim_{n\to \infty} \PP(T_{n,a} > c_{n,a}^P(\alpha)) =1,
\]
where $c_{n,a}^P(\alpha)$ denotes the $(1-\alpha)$-quantile  of the distribution of the permutation statistic $T_{n,a}^P$.
Since
\begin{equation}\label{konsist}
\liminf_{n\to \infty} \frac{T_{n,a}}{n} \ge \int_{\R^d} \left(\BE[\sin(t^\top  X)]\right)^2 \exp\left(-a \|t\|^2\right) \, \textrm{d} t
\end{equation}
almost surely (see display (5.1) of \cite{4}), we have
$\lim_{n \to \infty} T_{n,a} = \infty$ almost surely if the distribution of $X$ is not reflected symmetric. In view of the fact that
$c_{n,a}^P(\alpha)$ is bounded in probability almost surely, the test based on $T_{n,a}$ is consistent against any such alternative distribution.

To carry out the test in practice, one generates $M$ independent pseudo-random vectors $(U_1,\ldots,U_n)$, where $U_1,\ldots,U_n$ are
i.i.d. with a uniform distribution on $\{-1,+1\}$, and calculates the corresponding realizations $T_{n,a}^P(j)$, $1 \le j \le M$ (say), of the
permutation statistic $T_{n,a}^P$. The hypothesis $H_0$ is rejected at level $\alpha$, if the value of $T_{n,a}$ exceeds the empirical ($1-\alpha$)-quantile
of $T_{n,a}^P(j)$, $1 \le j \le M$. In Section \ref{secexam} we used $M=100000$, and the p-values given in Table 3 are based on $M=1000$ pseudo-random vectors.

\section{Behavior under fixed alternatives}\label{alter}
In this section, we assume that $\BE \|X\|^4 < \infty$ and that the distribution of $X$ is {\em not} symmetric. In view of affine invariance we further assume without loss of generality that
$\BE[X] =0$ and $\BE[XX^\top ] = \textrm{I}_d$, where I$_d$ stands for the unit matrix of order $d$. In what follows,
\begin{equation}\label{defrs}
\textrm{R}(t) = \BE\left[\cos(t^\top  X)\right], \quad  \textrm{I}(t) = \BE\left[\sin(t^\top X)\right], \quad t \in \R^d,
\end{equation}
denote the real and the imaginary part of the characteristic function of $X$, respectively.

The first result shows that the almost sure lower bound of $T_{n,a}/n$ figuring in (\ref{konsist}) is the stochastic limit
of $T_{n,a}/n$.

\begin{theorem}\label{thmlimalt} We have
\[
\frac{T_{n,a}}{n} \overset{\PP}{\longrightarrow} \Delta,
\]
where
\begin{equation}\label{defdelta}
\Delta = \int_{\R^d} \textnormal{I}(t)^2 \, \exp\left(-a\|t\|^2\right) {\textrm{\rm{d}}}t.
\end{equation}
\end{theorem}

\vspace*{5mm}
Interestingly, there is an alternative expression for the measure of distance $\Delta$ from symmetry figuring in (\ref{defdelta}).

\begin{theorem}\label{thmdeltadens}
We have
\begin{equation}\label{darstdelta}
\Delta = \frac{1}{4a^d} \int_{\R^d}
 \left(\BE\left[ \exp\left(- \frac{\|x-X\|^2}{2a}\right)-  \exp\left(- \frac{\|-x-X\|^2}{2a}\right)\right]  \right)^2 \, \textrm{{\rm d}} x.
\end{equation}
\end{theorem}

\vspace*{5mm}
To state a result on the limit distribution of $T_{n,a}$ under fixed alternatives, it will be convenient to introduce the
$\R^d$-valued functions
\begin{equation}\label{defcr}
\textnormal{C}(t) = \BE\left[X\cos\left(t^\top X\right)\right],  \quad \textnormal{S}(t) = \BE\left[X\sin\left(t^\top X\right)\right] \quad t \in \R^d.
\end{equation}

\begin{theorem}\label{thm3}
Under the assumptions stated at the beginning of this section, we have
\[
\sqrt{n}\left(\frac{T_{n,a}}{n}-\Delta\right)\overset{\mathcal{D}}{\longrightarrow} \textrm{{\rm N}}\left(0,\sigma^2\right),
\]
where
\begin{equation}\label{defsigma2}
\sigma^2=4\int_{\R^d}\int_{\R^d}K(s,t) \, \textnormal{I}(s) \textnormal{I}(t) \,
\exp\left(-a(\|s\|^2+\|t\|^2)\right)\, \textrm{{\rm d}}s\textrm{{\rm d}}t
\end{equation}
and
\begin{eqnarray*}
K(s,t)& = &\BE \left[\sin\left(s^\top X\right)\sin\left(t^\top X\right)\right]- \textnormal{I}(s) \textnormal{I}(t)
  -   \textnormal{R}(t) t^\top \textnormal{S}(s) -\textnormal{R}(s) s^\top \textnormal{S}(t)\\
 & &  +\textnormal{R}(s)  \textnormal{R}(t) s^\top t -\frac{1}{2} t^\top \BE\left[\sin\left(s^\top X\right) XX^\top \right] \textnormal{C}(t) + \frac{1}{2} \textnormal{I}(s) t^\top \textnormal{C}(t) \\
 & & -\frac{1}{2} s^\top \BE\left[\sin\left(t^\top X\right) XX^\top \right] \textnormal{C}(s) + \frac{1}{2} \textnormal{I}(t) s^\top \textnormal{C}(s)\\
 & & +\frac{1}{2} s^\top \textnormal{R}(s)\BE\left[Xt^\top XX^\top \right] \textnormal{C}(t)
 +\frac{1}{2} t^\top \textnormal{R}(t)\BE\left[ X s^\top XX^\top \right] \textnormal{C}(s)\\
& & +\frac{1}{4}\left\{ \textnormal{C}(s)^\top  \BE\left[XX^\top s t^\top X X^\top \right] \textnormal{C}(t) - s^\top \textnormal{C}(s) t^\top \textnormal{C}(t) \right\}, \quad s,t \in \R^d.
\end{eqnarray*}
\end{theorem}

{\sc Proof.} We use Theorem 1 of \cite{baebhe},  with  I$(t)$ corresponding to $z(t)$ in that paper. Putting
\[ W_n(t):=\frac{1}{\sqrt{n}}\sum_{j=1}^{n}\left(\sin\left(t^\top Y_{n,j}\right)- \textnormal{I}(t)\right), \quad t \in \R^d,
\]
we will show that $W_n(\cdot) \overset{\mathcal{D}}{\longrightarrow} W(\cdot)$ in $\mathcal{L}^2$, where $W(\cdot)$ is a centred
Gaussian random element of $\mathcal{L}^2$ having covariance kernel $K(s,t)$ figuring in the statement of Theorem \ref{thm3}.
Denoting by $\langle \cdot,\cdot \rangle$ the inner product in $\mathcal{L}^2$ and observing that, with $\textnormal{I}(\cdot)$ defined in
(\ref{defrs}),
\[
\sqrt{n}\left(\frac{T_{n,a}}{n}-\Delta\right) = 2\langle W_n,\textnormal{I}\rangle + \frac{1}{\sqrt{n}} \|W_n\|_{\mathcal{ L}^2}^2,
\]
the continuous mapping theorem yields $\langle W_n,\textnormal{I}\rangle \overset{\mathcal{D}}{\longrightarrow} \langle W, \textnormal{I}\rangle$
as well as  $\|W_n\|^2_{\mathcal{L}^2} \overset{\mathcal{D}}{\longrightarrow} \|W\|^2_{\mathcal{L}^2}$, whence
\[
\sqrt{n}\left(\frac{T_{n,a}}{n}-\Delta\right) \overset{\mathcal{D}}{\longrightarrow} 2\langle W,\textnormal{I}\rangle.
\]
The distribution of $2\langle W,\textnormal{I}\rangle$ is the required normal distribution N$(0,\sigma^2)$. The proof of
 $W_n(\cdot) \overset{\mathcal{D}}{\longrightarrow} W(\cdot)$ will only be sketched since it closely parallels the proof of
 Theorem 3.1 of \cite{4}. Let
 \begin{eqnarray*}
\overline{W}_n(t)&:= & \frac{1}{\sqrt{n}}\sum_{j=1}^{n}\left(\sin\left(t^\top X_j\right)+t^\top \Delta_{n,j}\cos(t^\top X_j)-\textnormal{I}(t)\right), \\
W_n^*(t) &:=&\frac{1}{\sqrt{n}}\sum_{j=1}^{n}\bigg(\sin\left(t^\top X_j\right)- \textnormal{I}(t) - t^\top \textnormal{R}(t)X_j-\frac{1}{2}t^\top \left(X_jX_j^\top -\textnormal{I}_d\right) \textnormal{C}(t)\bigg),
\end{eqnarray*}
where $\Delta_{n,j}$ is given in (\ref{delnj}).
Since $W_n = (W_n - \overline{W}_n) + (\overline{W}_n - W_n^*) + W_n^*$, the main steps of the proof are to show
$\left\|W_n-\overline{W}_n\right\|_{\mathcal{L}^2}=o_\PP(1)$,
$\left\|\overline{W}_n-W_n^*\right\|_{\mathcal{L}^2}=o_\PP(1)$ and $W_n^*\ \overset{\mathcal{D}}{\longrightarrow} W$  in $\mathcal{L}^2$.
The details are omitted. Notice that the convergence $W_n^*\ \overset{\mathcal{D}}{\longrightarrow} W$ follows from the Lindeberg--L\'{e}vy type central limit theorem
in separable Hilbert spaces (see, e.g., \cite{bosq}) since the summands comprising  $W_n^*$ are i.i.d. centred random elements of $\mathcal{L}^2$. \qed

\section{Estimation of $\sigma^2$}\label{secesti}
Theorem \ref{thm3} paves the way to an asymptotic confidence interval for $\Delta$ provided that a consistent estimator $\widehat{\sigma}_n^2 =
\widehat{\sigma}_n^2(X_1,\ldots,X_n)$ of the variance $\sigma^2$ figuring in (\ref{defsigma2}) is available. Since Theorem \ref{thm3} requires
$\BE (X) =0$ and $\BE(XX^\top ) = \textnormal{I}_d$, we base such an estimator on the empirically standardized data
defined in (\ref{defynj}), where we put $Y_j = Y_{n,j}$ for the sake of brevity in what follows. Moreover, let $w(s,t) =  \exp\left(-a(\|s\|^2+\|t\|^2)\right)$. Such an estimator is
\begin{equation}\label{defsigmandach}
\widehat{\sigma}_n^2 = 4 \int_{\R^d}\int_{\R^d}K_n(s,t) \, \textnormal{I}_n(s) \textnormal{I}_n(t) \,
w(s,t)\, \textrm{{\rm d}}s\textrm{{\rm d}}t.
\end{equation}
Here, $K_n(s,t)$ is the empirical version of $K(s,t)$ figuring in the statement of Theorem \ref{thm3}. This version originates from $K(s,t)$
by replacing the functions $\textnormal{R}(\cdot)$, $\textnormal{I}(\cdot)$, $\textnormal{C}(\cdot)$ and $\textnormal{S}(\cdot)$ defined in
(\ref{defrs}) and (\ref{defcr}) with their respective empirical counterparts
\begin{eqnarray*}
\textnormal{R}_n(t) & = & \frac{1}{n} \sum_{j=1}^n \cos\left(t^\top Y_j\right), \quad
\textnormal{I}_n(t) = \frac{1}{n} \sum_{j=1}^n \sin\left(t^\top Y_j\right),\\
\textnormal{C}_n(t) & = & \frac{1}{n} \sum_{j=1}^n Y_j\cos\left(t^\top Y_j\right), \quad \textnormal{S}_n(t)
 = \frac{1}{n} \sum_{j=1}^n Y_j \sin\left(t^\top Y_j\right), \ \ t \in \R^d,
\end{eqnarray*}
and doing the same with each of the  five explicitly designated expectations figuring in the definition of $K(s,t)$. Thus, for example,
$\BE \left[\sin\left(s^\top X\right)\sin\left(t^\top X\right)\right]$ is replaced with $n^{-1}\sum_{j=1}^n \sin(s^\top Y_j)\sin(t^\top Y_j)$ etc.
To give an expression of $\widehat{\sigma}_n^2$ that does not involve any integration an is thus amenable to computational purposes, we put
\begin{eqnarray*}
\rho_1(u,v) & := & \int_{\R^d} \sin(u^\top t)\sin (v^\top t) \exp(-a\|t\|^2)\, \textnormal{d}t,\\ \label{intr2}
\rho_2(u,v) & := & \int_{\R^d} t \cos(u^\top t)\sin (v^\top t) \exp(-a\|t\|^2)\, \textnormal{d}t, \quad u,v \in \R^d.
\end{eqnarray*}
These integrals can be evaluated to give
\begin{eqnarray*}
\rho_1(u,v) \! & \! = \! & \! \frac{1}{2} \left(\frac{\pi}{a}\right)^{d/2}
\left(\exp\left( - \frac{\|u-v\|^2}{4a}\right) -  \exp\left(- \frac{\|u+v\|^2}{4a}\right)\right),\\
\rho_2(u,v) \! & \! = \! & \! \frac{1}{4a} \left(\frac{\pi}{a}\right)^{d/2} \! \left(\!  (v-u)\exp \! \left(\! - \frac{\|v-u\|^2}{4a}\right)+
(v+u)\exp\! \left(\! - \frac{\|v+u\|^2}{4a}\right)\! \right).
\end{eqnarray*}
Notice that the function $\rho_2$ takes values in $\R^d$. Letting
\begin{eqnarray*}
V_{n,r} & := & \frac{1}{n^2}\sum_{i,j=1}^n \rho_r(Y_i,Y_j), \quad \overline{V}_{n,r}(y) := \frac{1}{n}\sum_{\ell =1}^n \rho_r(y,Y_\ell ), \quad y \in \R^d,\
 r \in \{1,2\},\\
\Sigma_n & := & \frac{1}{n^2} \sum_{i,k=1}^n \rho_1(Y_i,Y_k)Y_iY_i^\top, \quad  \Gamma_n := \frac{1}{n^2} \sum_{i,\ell =1}^n \rho_2(Y_i,Y_\ell) Y_i^\top,
\end{eqnarray*}
a computationally feasible expression for $\widehat{\sigma}_n^2$ is given as follows.

\begin{proposition}\label{proposition1}
We have
\begin{eqnarray*}
\widehat{\sigma}_n^2 & = & \frac{4}{n} \sum_{j=1}^n \overline{V}_{n,1}(Y_j)^2  - 4 V_{n,1}^2
 - 8 \left(\! \frac{1}{n} \sum_{j=1}^n \overline{V}_{n,1}(Y_j) Y_j \! \right)^\top \! V_{n,2}
+4 \big{\|} V_{n,2}\big{\|^2}\\
& & - 4 \, \textnormal{tr} \! \left(\! \Sigma_n \left( \frac{1}{n} \! \sum_{j=1}^n \overline{V}_{n,2}(Y_j) Y_j^\top \! \right) \! \right) + 4
 V_{n,1} \, \frac{1}{n} \sum_{j=1}^n Y_j^\top \overline{V}_{n,2}(Y_j)\\
  & & + 4 V_{n,2}^\top \left(\frac{1}{n^2}\sum_{j,k=1}^n Y_jY_j^\top Y_kY_j^\top \overline{V}_{n,2}(Y_k)\right) \\
  & & + \frac{1}{n} \sum_{j=1}^n \left(Y_j^\top \Gamma_n Y_j\right)^2 - \left( \frac{1}{n} \sum_{i=1}^n Y_i^\top \overline{V}_{n,2}(Y_i)\right)^2.
\end{eqnarray*}
\end{proposition}

\vspace*{3mm}
The next result shows that $\widehat{\sigma}_n^2$ defined in \eqref{defsigmandach} is a consistent estimator of
 $\sigma^2$ defined in \eqref{defsigmandach}.

\vspace*{3mm}
\begin{theorem}\label{thmconsist}
Under the standing assumptions, we have
\[
\widehat{\sigma}_n^2 \overset{\PP}{\longrightarrow} \sigma^2.
\]
\end{theorem}

The proof is extremely tedious but in principle straightforward. A similar problem was encountered in
\cite{gu} in the context of estimating the variance  of the limit normal distribution of the BHEP test for multivariate normality
under a fixed alternative distribution. Details are given in Section \ref{secproofs}.

\section{Example}\label{secexam}
Suppose that $\PP^X$ is the normal mixture
\[
X \overset{\mathcal{D}}{=} TY_1+ (1-T)Y_2,
\]
where $T,Y_1,Y_2$ are independent, $\PP(T=1) = p = 1- \PP(T=0)$, $0 \le p < 1$, $Y_1 \overset{\mathcal{D}}{=} \textnormal{N}(\textnormal{e}_1,\textnormal{I}_d-\frac{p}{1-p}\textnormal{e}_1\textnormal{e}_1^\top)$ and
$Y_2 \overset{\mathcal{D}}{=} \textnormal{N}(-p/(1-p) \textnormal{e}_1,\textnormal{I}_d-\frac{p}{1-p}\textnormal{e}_1\textnormal{e}_1^\top)$, where $\textnormal{e}_1 = (1,0,\ldots,0)^\top $ is the first canonical
unit vector in $\R^d$. In view of $T^2=T$ and independence, we have
\begin{eqnarray*}
\BE(X) & = & \BE(T) \BE(Y_1) + (1-\BE(T))\BE(Y_2) = 0,\\
\BE(XX^\top ) & = & \BE(T) \BE(Y_1Y_1^\top ) + (1-\BE(T))\BE(Y_2Y_2^\top ) = \textnormal{I}_d.
\end{eqnarray*}
The addition Theorem for the sine function gives
\[
\sin(t^\top X) \overset{\mathcal{D}}{=}  \sin(Tt^\top Y_1) \cos((1-T)t^\top Y_2) + \cos(Tt^\top Y_1) \sin((1-T)t^\top Y_2),
\]
and conditioning on $T$ it follows that
\[
\textnormal{I}(t)  =  \BE[\sin(t^\top X)]  =  p \, \BE[\sin(t^\top Y_1)] + (1-p) \, \BE[\sin(t^\top Y_2)].
\]
Writing $t=(t_1,\ldots,t_d)^\top $, we have \begin{align*}t^\top Y_1 \overset{\mathcal{D}}{=}&\  \textnormal{N}\left(t_1,\|t\|^2+\left(\frac{1-2p}{1-p}-1\right)t_1^2\right),\\ t^\top Y_2 \overset{\mathcal{D}}{=}&\  \textnormal{N}\left(-pt_1/(1-p),\|t\|^2+\left(\frac{1-2p}{1-p}-1\right)t_1^2\right).\end{align*} Since the characteristic function of the
normal distribution N$(\mu,\sigma^2)$ is $\exp(\textnormal{i}\xi \mu - \sigma^2 \xi^2/2)$, $\xi \in \R$,
it follows that
\[
\textnormal{I}(t) = \exp\left(- \frac{\|t\|^2+\left(\frac{1-2p}{1-p}-1\right)t_1^2}{2}\right) \left(p \sin t_1 - (1-p) \sin\left(\frac{pt_1}{1-p} \right)\right).
\]
Thus
\begin{eqnarray*}
\Delta & = & p^2 \int_{\R^d} \sin^2(t_1) \exp\left(-(1+a)\|t\|^2-\left(\frac{1-2p}{1-p}-1\right)t_1^2\right) \textnormal{d}t \\
& & \quad + (1-p)^2 \int_{\R^d} \sin^2\left(\frac{pt_1}{1-p}\right)  \exp\left(-(1+a)\|t\|^2-\left(\frac{1-2p}{1-p}-1\right)t_1^2\right) \textnormal{d}t \\
& & \quad - 2p(1-p) \int_{\R^d} \sin(t_1)\,  \sin\left(\frac{pt_1}{1-p}\right)  \exp\left(-(1+a)\|t\|^2-\left(\frac{1-2p}{1-p}-1\right)t_1^2\right) \textnormal{d}t.
\end{eqnarray*}
Since $\int_{-\infty}^\infty \exp(-(1+a)\xi^2) \, \textnormal{d}\xi = \sqrt{\pi/(1+a)}$, the computation of $\Delta$ boils down to
the calculation of integrals of the type
\[
 \int_{-\infty}^\infty \sin(\alpha \xi)\sin(\beta \xi) \exp(-\gamma \xi^2) \textnormal{d}\xi
 =  \frac{\sqrt{\pi}}{2\sqrt{\gamma}} \left( \exp\left(- \frac{(\alpha - \beta)^2}{4 \gamma}\right) - \exp\left(- \frac{(\alpha+\beta)^2}{4 \gamma}\right) \right),
 \]
  where $\alpha, \beta \in \R$ and $\gamma >0$. After tedious but straightforward calculations, one obtains
  \begin{eqnarray*}
  \Delta \! & \! = \! & \! \left(\frac{\pi}{a+1}\right)^{(d-1)/2} \! \! \sqrt{\frac{{\pi}}{\gamma_a}} \left[ \frac{p^2}{2} \! \left(\! 1- \exp\left(- \frac{1}{\gamma_a}\right)\right) + \frac{(1-p)^2}{2} \!
  \left(\! 1- \exp\left(\! - \frac{p^2}{(1-p)^2\gamma_a} \! \right) \! \right) \right. \\
  \! & \! = \! & \!  \qquad \left. - p(1-p)\left(\exp\left(- \frac{(1-2p)^2}{4(1-p)^2\gamma_a} \right) - \exp \left(- \frac{1}{4(1-p)^2\gamma_a} \right)\right) \right],
  \end{eqnarray*}
where $\gamma_a=a+(1-2p)/(1-p)$.

Using the above normal mixture with $p=0.25$ and $p=0.4$ in the case $d=1$, we investigated whether the estimator  $T_{n,a}/n$ of $\Delta$
is useful for practical purposes. Since the normal mixture exhibits  fairly weak asymmetry, we studied the performance of $T_{n,a}/n$ also
on centered Exp(1) distributed samples, which represent a much stronger degree of asymmetry. To obtain
 a reasonable conclusion, we computed the underlying values of $\Delta$ and $\sigma^2$ for the latter distribution by means of  numerical integration.

Regarding the choice of the parameter $a$, note that small values of $a$ implicate bigger values for both $\Delta$ and $T_{n,a}/n$,  and likewise for
$\sigma^2$ and $\widehat{\sigma}_n^2$. To bypass computational inaccuracies and to avoid negative values of $\widehat{\sigma}_n^2$ that sometimes show up in
small sample sizes, we used mainly small values for $a$, which seems  to have no disadvantages at all. Nevertheless,
the qualitative behavior of the estimates is similar if the sample size is big enough. To see the effect of $a$, the outcome of the simulation study is displayed in Table 1
for the case $a=0.01$ and in Table 2 for the case $a=0.1$. In these tables, the normal mixtures with $p=0.25$ and $p=0.4$ are denoted by N1 and N2, respectively, and the centered standard exponential distribution is denoted by E.  For each combination of the sample size $n$, the parameter $a$ and the underlying distribution, we performed 1000
simulations and computed the sample mean of $T_{n,a}/n$ (denoted by $\varnothing T_{n,a}/n$) and the sample variance $\widehat \sigma_n^2$ (denoted by $\varnothing\widehat{\sigma}_n^2$)
as estimates of $\Delta$ and $\sigma^2$, respectively. Thereby we calculated an approximation for the $95\%$ confidence interval and observed how often the interval contained $\Delta.$ The average number per 100 samples can be seen in the columns called "estimated $ \alpha "$. Furthermore, we noted the total number of negative estimates for $\sigma ^2$ as $"\widehat\sigma_n^2<0"$ and the relative mean squared error of  $T_{n,a}/n$, i.e. $$\frac{\frac{1}{1000}\sum_{j=1}^{1000}(\frac{T_n}{n}-\Delta)}{\Delta},$$ as "relative MSE". In Table 1, the true values of $\Delta$ are  $\Delta=0.01039$ for N1, $\Delta= 0.05062$ for N2 and $\Delta= 0.55771$ for E. Furthermore, the value of $\sigma^2$ is  $3.0409$ for the distribution E.

\begin{table}[htb]
	\begin{center}
		\begin{tabular}{c l| c|c|c|c|c}
			\hspace*{0.1cm}&\hspace*{0.1cm} $n$&\hspace*{0.1cm} $\varnothing T_{n,a}/n$\hspace*{0.1cm} & $\varnothing\widehat{\sigma}_n^2$ &\hspace*{0.1cm} $\widehat\sigma_n^2<0$\hspace*{0.1cm} & \hspace*{0.1cm}estimated $\alpha$\hspace*{0.1cm} &\hspace*{0.1cm}relative MSE  \hspace*{0.1cm} \\ \hline \hline
			N1 &40& 0.1945 & 1.4770 & 1 & 93.7 & 4.3572 \\
			& 80& 0.1022&0.6431&1&96.6& 1.0978\\
			&100& 0.0846 & 0.4946 & 0 & 96.3 &0.7377 \\
			&250& 0.0391& 0.1986 & 0 & 97.9 &0.1221\\
			&500& 0.0240& 0.1211 & 0&  98.0 & 0.0344\\ \hline
			N2 &40& 0.2355 & 1.9090 & 1 & 97.0 &1.0283 \\
			& 80& 0.1465&1.0602&0&98.4& 0.3180\\
			&100& 0.1300 & 0.9362 & 0 & 98.2 &0.2428 \\
			&250& 0.0798& 0.6020 & 0 & 97.7 &0.0579\\
			&500& 0.0641& 0.5073 & 0&  95.9 &  0.0218\\ \hline
			
			E &40& 0.7120 & 6.2629 & 0 & 98.4 & 0.2087 \\
			& 80& 0.6442&5.1665&0&97.2&0.1095 \\
			&100& 0.6244 & 4.9267 & 0 & 97.7 &0.0771 \\
			&250& 0.5866& 4.5024 & 0 & 97.2 &0.0288\\
			&500& 0.5726 &4.3538 & 0&  96.8 & 0.0134\\

		\end{tabular}
	\end{center}
\end{table}
\vspace*{-7mm}
\begin{center}
{\bf Table 1:} 	Estimated values based on $1000$ samples of the distributions N1, N2 and E, $a=0.01$.
\end{center}

In Table 2, the true values of $\Delta$ are $\Delta=0.00713$ for N1, $\Delta= 0.02889$ for N2  and $\Delta= 0.29080$ for the centered standard exponential distribution E. For the latter distribution, the value of $\sigma^2$ is $\sigma^2=0.8875$.

As each table indicates, the desired properties can also be seen in practical applications. Even for small sample sizes the computed intervals maintain the nominal level, and the estimator $T_{n,a}/n$ quantifies the departure from symmetry for fixed $a$. Furthermore, the relative mean squared error decreases drastically as the sample size increases.

\clearpage

\begin{table}[htb]
	\begin{center}
		\begin{tabular}{c l| c|c|c|c|c}
			\hspace*{0.1cm}&\hspace*{0.1cm} $n$&\hspace*{0.1cm} $\varnothing T_{n,a}/n$\hspace*{0.1cm} & $\varnothing\widehat{\sigma}_n^2$ &\hspace*{0.1cm} $\widehat\sigma_n^2<0$\hspace*{0.1cm} & \hspace*{0.1cm}estimated $\alpha$\hspace*{0.1cm} &\hspace*{0.1cm}relative MSE  \hspace*{0.1cm} \\ \hline \hline
			N1 &40& 0.0436 & 0.1452 & 14 & 95.6 &0.4535 \\
			& 80& 0.0240&0.0800&6&98.0& 0.1104\\
			&100& 0.0220 & 0.0733 & 3 & 98.0 & 0.0895 \\
			&250& 0.0130& 0.0422 & 2 & 95.3 &0.0213\\
			&500& 0.0100& 0.0327 & 1&  91.5 &0.0087 \\ \hline
			N2 &40&0.0614 & 0.2721 & 24 & 95.0 & 0.1897 \\
			& 80& 0.0488&0.2086&11&92.1& 0.0791\\
			&100& 0.0475 & 0.2066 & 3 & 91.9 & 0.0694 \\
			&250& 0.0344& 0.1554 & 1 & 92.1 & 0.0188\\
			&500& 0.0321& 0.1491 & 0&  93.2 & 0.0153\\ \hline

			E &40& 0.3052 & 0.9429 & 1 & 91.9 & 0.0190 \\
			& 80& 0.3026&0.9135&0&94.0& 0.0113\\
			&100& 0.2994 & 0.9028 & 0 & 94.7 &0.0084 \\
			&250& 0.2949& 0.8679 & 0 & 95.5 &0.0033\\
			&500& 0.0294& 0.8541 & 0&  95.2 & 0.0017\\
		
		\end{tabular}
\end{center}
\end{table}
\vspace*{-7mm}
\begin{center}
	{\bf Table 2:} Estimated values based on $1000$ samples of the distributions N1, N2 and E, $a=0.1$.
	\end{center}

\section{A real data example}\label{secrealdata}
We consider a data set that originated from a health survey of paint sprayers in a car assembly plant. This data set, which is given in
  \cite{Royston}, contains 103 observations, each consisting of 6 variates, namely:

 1. haemoglobin concentration,

  2. PCV packed cell volume,

  3. white blood cell count,

  4. lymphocyte count,

  5. neutrophil count,

  6. serum lead concentration.

As is a common procedure for haematological data (see e.g., \cite{Royston}), we applied a logarithmic transformation to each of the variates  3. - 6., since these exhibit skewed distributions. Royston \cite{Royston} first investigated whether the transformed data arises from a normal distribution. Since three observations seem to be outliers, they were removed. By applying a multivariate generalization of the Shapiro--Wilk test for univariate normality, Royston deduced that the 6-dimensional data showed significant departures from normality, although such a conclusion could not be drawn for any of the bivariate marginal distributions.

\begin{figure}[htb]
	\begin{center}
\includegraphics[width=0.9\textwidth]{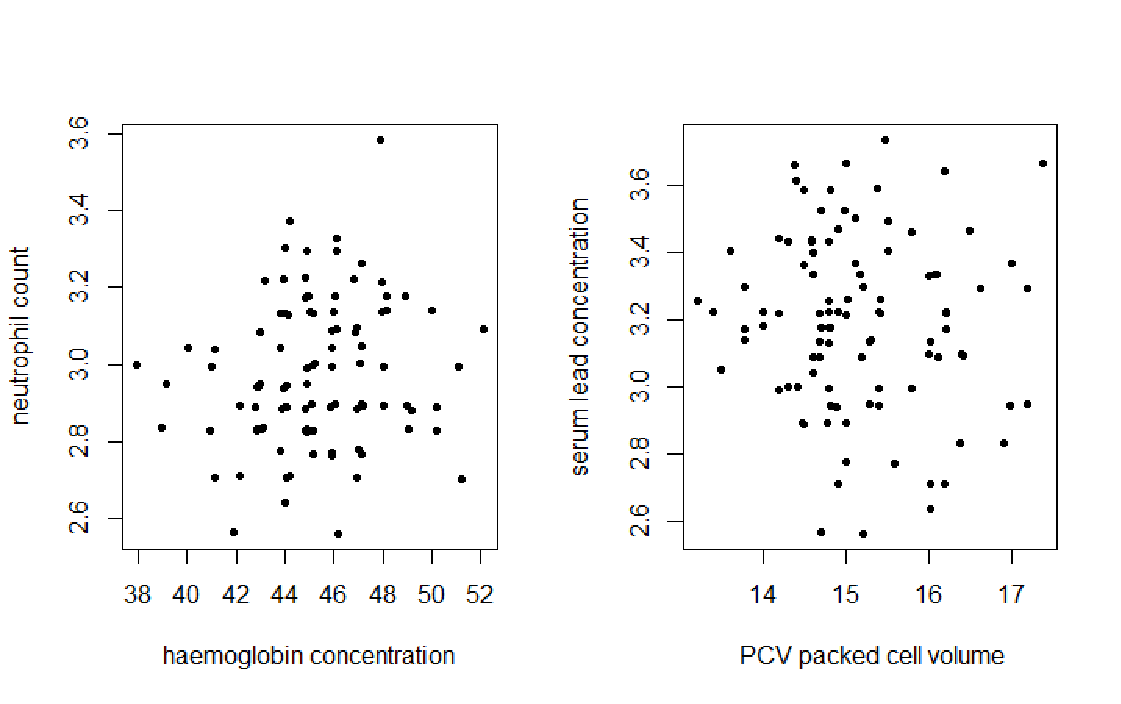}
\end{center}
\begin{center}
{\bf Fig.1:} Scatterplots  haemoglobin concentration - neutrophil count (left)\\ and  PCV packed cell volume - serum lead concentration (right)
\end{center}
\end{figure}

From an application of a covariance-matrix based Wald-test to the transformed full data set, Schott \cite{Schott} arrived at the same result.
Since a test for elliptical symmetry, applied to the same data set, gave a p-value of 0.11, Schott argued that it is not unreasonable to assume that
the sample originates from an elliptical distribution.

Using a Chi-square type statistic for testing for elliptical symmetry, Batsidis et al. \cite{Batsidis} even obtained a p-value larger than 0.9 and thus did not find any evidence for rejecting the hypothesis of elliptical symmetry.  The latter findings are in stark contrast to the results that originate when applying the
HKM-test to the full data set. Astonishingly, the test rejected the hypothesis of {\em central symmetry} with a p-value of $7\cdot10^{-5}$ using $a=1$. Taking $a=0.5$, $a=2$ and $a=4$ leads to p-values of a  similar magnitude. Since central symmetry is a necessary condition for elliptical symmetry, we can also strongly reject the hypothesis of elliptical symmetry of the 6-variate full data set.

To investigate whether the declared outliers are responsible for rejecting symmetry, we removed these
 outliers (observations 21, 47 and 52 in the data set given in \cite{Royston})  and applied the HKM-test. Again taking $a=0.5, a=1, a=2$ and $a=4$, we obtained p-values of magnitude $10^{-3}.$ Consequently, also the remaining data exhibit strong asymmetry.

We finally addressed the question whether any bivariate combination of the 6-dimensional logarithmically transformed data (without outliers) is compatible with the hypothesis of reflected symmetry. Looking at the two plots in Figure 1, both combinations seem to be equally symmetric or rather skew. However taking $a=1$ we obtained the p-values given in Table 3. Apparently, the desired 5\% level of significance is only exeeded for the combinations `haemoglobin concentration - white blood cell count' and `haemoglobin concentration - neutrophil count'. Consequently there is no evidence of departure from symmetry for the right-hand combination in Figure 1, whereas the left-hand one is certainly skew.

\clearpage

\begin{table}[h]
\begin{center}
	\begin{tabular}{|c l| c|c|c|c|c|}\hline
		\hspace*{0.1cm}&\hspace*{0.1cm}&\hspace*{0.1cm} p.c.v.\hspace*{0.1cm} & w.b.c. count &\hspace*{0.1cm} l. count\hspace*{0.1cm} & \hspace*{0.1cm}n. count\hspace*{0.1cm} &\hspace*{0.1cm}s.l. con. \hspace*{0.1cm} \\ \hline
		&haem. con. &0.158 &0.027 & 0.227 & 0.028 & 0.111\\
		&p.c.v.& &  0.252 & 0.694 & 0.531 & 0.699\\
		&w.b.c. count&  & & 0.164 & 0.076 & 0.286 \\
		&l. count  & & & & 0.732 & 0.381\\
		& n. count & & & & & 0.645 \\ \hline
	
	\end{tabular}
\end{center}
	\label{tab:meinetabelle}
\end{table}
\vspace*{-5mm}
\begin{center}
{\bf Table 3:} p-values of the bivariate HKM-test with parameter $a=1$
\end{center}

\section{Proofs}\label{secproofs}

{\sc Proof of Theorem \ref{thmlimalt}.}  Putting
\[
\widetilde{T}_{n,a} = \int_{\R^d} \left(\frac{1}{\sqrt{n}}\sum_{j=1}^{n}\sin\left(t^\top X_j\right)\right)^2\exp\left(-a\|t\|^2\right)\textrm{d}t,
\]
we have
\[
\frac{\widetilde{T}_{n,a}}{n}=\left\|\frac{1}{{n}}\sum_{j=1}^{n}\sin\left(\bullet^\top X_j\right)\right\|_{\mathcal{L}^2}^2,
\]
where $\|\cdot \|_{\mathcal{L}^2}$ denotes the norm in $\mathcal{L}^2$.
The strong law of large numbers in Banach spaces yields
\[ \lim_{n\to \infty}
\left\|\frac{1}{n}\sum_{j=1}^{n}\sin\left(\bullet^\top X_j\right)-\BE\left[\sin\left(\bullet^\top X\right)\right]\right\|_{\mathcal{L}^2}^2 = 0
\]
$\PP$-almost surely,
whence
\begin{equation}\label{ttildeconv}
\lim_{n\to \infty} \frac{\widetilde{T}_{n,a}}{n}  = \Delta
\end{equation}
$\PP$-almost surely. Since
\[
\left|\frac{1}{{n}}\sum_{j=1}^{n}\sin\left(t^\top X_j\right)+\frac{1}{{n}}\sum_{j=1}^{n}\sin\left(t^\top Y_{n,j}\right)\right|\leq 2,
\]
it follows that
\[
\frac{|\widetilde{T}_{n,a}- T_{n,a}|}{n}\leq 2 \int_{\R^d} \left|\frac{1}{n}\sum_{j=1}^{n}\sin\left(t^\top X_j\right)-\frac{1}{n}\sum_{j=1}^{n}\sin\left(t^\top Y_{n,j}\right)\right|\exp\left(-a\|t\|^2\right)\textrm{d}t.
\]
Putting
\begin{equation}\label{delnj}
\Delta_{n,j} = Y_{n,j} - X_j = S_n^{-1/2}(X_j- \overline{X}_n) - X_j,
\end{equation}
the inequalities $|\sin a - \sin b| \le |a-b|$ and $|t^\top z| \le \|t\| \cdot \|z\|$ give
\[
\frac{|\widetilde{T}_{n,a}- T_{n,a}|}{n}\leq 2 \left(\frac{1}{n} \sum_{j=1}^n \|\Delta_{n,j}\|\right) \int_{\R^d} \|t\|  \ \exp(-a\|t\|^2) \, \textrm{d} t.
\]
Writing $\textnormal{tr}(A)$ for the trace of a square matrix $A$, we have
\begin{eqnarray*}
\frac{1}{\sqrt{n}} \sum_{j=1}^n \|\Delta_{n,j}\|^2 & = & \sqrt{n} \, \textnormal{tr}  \left((S_n^{-1/2}- \textnormal{I}_d)^2 \frac{1}{n} \sum_{j=1}^n X_jX_j^\top \right)\\
& & \quad - 2 \overline{X}_n^\top S_n^{-1/2} \sqrt{n}(S_n^{-1/2}-\textnormal{I}_d)\overline{X}_n + \sqrt{n} \overline{X}_n^\top S_n^{-1} \overline{X}_n
\end{eqnarray*}
and
\[
\sqrt{n}(S_n^{-1/2}-\textnormal{I}_d) = - \frac{1}{2\sqrt{n}}\sum_{j=1}^n (X_jX_j^\top- \textnormal{I}_d) + O_\PP(n^{-1/2}),
\]
see p. 9 of \cite{hewa}. Since $\BE \|X\|^4<\infty$ implies $\sqrt{n}(S_n^{-1/2}-\textnormal{I}_d) = O_\PP(1)$, it follows that
\begin{equation}\label{deltabound}
\frac{1}{\sqrt{n}} \sum_{j=1}^n \|\Delta_{n,j}\|^2 = o_\PP(1).
\end{equation}
In view  of (\ref{ttildeconv}) and  the Cauchy-Schwarz estimate
\[
\frac{1}{n} \sum_{j=1}^n \|\Delta_{n,j}\| \le \left(\frac{1}{n} \sum_{j=1}^n \|\Delta_{n,j}\|^2\right)^{1/2},
\]
we have
\begin{equation}\label{deltanquer}
\frac{1}{n} \sum_{j=1}^n \|\Delta_{n,j}\| = o_\PP(1),
\end{equation}
and the proof is completed. \qed

\vspace*{4mm}

{\sc Proof of Theorem \ref{thmdeltadens}.} Denote the right-hand side of (\ref{darstdelta}) by $\widetilde{\Delta}$. From (\ref{tna3}) we have
\[
\frac{T_{n,a}}{n} = \frac{(2\pi)^d}{4}\int_{\R^d}\left(\widehat{f}_{n,a}(x)-\widehat{f}_{n,a}(-x)\right)^2\textrm{d}x.
\]
 We show
$\lim_{n\to \infty} \BE[T_{n,a}/n] = \widetilde{\Delta}$ and  $\lim_{n\to \infty} \BV(T_{n,a}/n) =0$.
Since a constant stochastic limit is uniquely determined, the assertion follows. Fubini's theorem gives
\[
\BE\left[ \frac{T_{n,a}}{n}\right]  = \frac{(2\pi)^d}{4}\int_{\R^d} \BE \left[
\left(\widehat{f}_{n,a}(x)-\widehat{f}_{n,a}(-x)\right)^2\right] \textrm{d}x.
\]
Using (\ref{kerneld}) and  expanding the round bracket, we obtain
\[
\widehat{f}_{n,a}(x)^2 = \frac{1}{(2\pi a)^d} \, \frac{1}{n^2} \, \sum_{i,j=1}^n \exp\left(-\frac{\|x-Y_{n,i}\|^2}{2a}\right)
\exp\left(-\frac{\|x-Y_{n,j}\|^2}{2a}\right).
\]
Taking expectations,  symmetry arguments, the inequality $\exp(-\xi) \le 1$, $\xi \ge 0$, almost sure convergence of
$Y_{n,j}$ to $X_j$ for fixed $j$, dominated convergence and independence yield
\[
\lim_{n\to \infty} \BE \left[\widehat{f}_{n,a}(x)^2\right] =  \frac{1}{(2\pi a)^d} \, \BE\left[\exp \left(- \frac{\|x-X\|^2}{2a}\right)\right]^2.
\]
The other terms are treated similarly, and thus $\lim_{n\to \infty} \BE[T_{n,a}/n] = \widetilde{\Delta}$.
To prove $\lim_{n\to \infty}$ $\BV(T_{n,a}/n) =0$, start with
\[
\left(\frac{T_{n,a}}{n}\right)^2 = \int_{\R^d}\int_{\R^d} \left(\widehat{f}_{n,a}(x)-\widehat{f}_{n,a}(-x)\right)^2 \left(\widehat{f}_{n,a}(y)-\widehat{f}_{n,a}(-y)\right)^2 \textrm{d}x \, \textrm{d}y
\]
and use the techniques indicated above to show that $\lim_{n\to \infty} \BE[(T_{n,a}/n)^2] = \widetilde{\Delta}^2$.
Hence $\lim_{n\to \infty} \BV(T_{n,a}/n) =0$, and the assertion follows. \qed

\vspace*{3mm}

{\sc Proof of Proposition \ref{proposition1}.} Starting with \eqref{defsigmandach}, the proof follows from straightforward but tedious calculations and symmetry arguments using
\begin{eqnarray*}
 \iint  \frac{1}{n} \sum_j \sin(s^\top Y_j)\sin(t^\top Y_j) \, \textnormal{I}_n(s) \textnormal{I}_n(t) \,
w(s,t)\, \textrm{{\rm d}}s\textrm{{\rm d}}t  \! & \! = \! &  \!   \frac{1}{n^3} \! \sum_{j,k,\ell} \! \rho_1(Y_j,Y_k) \rho_1(Y_j,Y_\ell)\\
\! & \! = \! & \! \frac{1}{n} \sum_j \overline{V}_{n,1}(Y_j)^2,
\end{eqnarray*}
\[
\iint  \textnormal{I}_n^2(s) \textnormal{I}^2_n(t)
w(s,t)\, \textrm{{\rm d}}s\textrm{{\rm d}}t = \left(\frac{1}{n^2} \sum_{j,k} \rho_1(Y_j,Y_k)\right)^2 = V_{n,1}^2,
\]
\begin{eqnarray*}
\iint  \! \textnormal{R}_n(t) t^\top \textnormal{S}_n(s) \textnormal{I}_n(s)\textnormal{I}_n(t) w(s,t) \,  \textrm{{\rm d}}s\textrm{{\rm d}}t
\! & \! = \! & \!    \left(\! \frac{1}{n^2} \sum_{j,k} \rho_1(Y_j,Y_k) Y_j \! \right)^\top \! \! \left(\! \! \frac{1}{n^2} \! \sum_{i,\ell}  \rho_2(Y_i,Y_\ell)\! \right)\\
\! & \! = \! &  \!  \left(\! \frac{1}{n} \sum_j \overline{V}_{n,1}(Y_j) Y_j \! \right)^\top \! V_{n,2},
\end{eqnarray*}
\[
\iint  \textnormal{R}_n(s) \textnormal{R}_n(t) s^\top t \, \textnormal{I}_n(s)\textnormal{I}_n(t) w(s,t) \,  \textrm{{\rm d}}s\textrm{{\rm d}}t
= \Big{\|} \frac{1}{n^2} \sum_{i,k} \rho_2(Y_i,Y_k)\Big{\|^2} = \|V_{n,2}\|^2,
\]
\begin{eqnarray*}
& &  \iint  t^\top \frac{1}{n} \sum_i \sin(s^\top Y_i) Y_iY_i^\top \textnormal{C}_n(t)
  \textnormal{I}_n(s) \textnormal{I}_n(t) \,
w(s,t)\, \textrm{{\rm d}}s\textrm{{\rm d}}t \\
& = & \frac{1}{n^4} \sum_{i,j,k,\ell} \rho_1(Y_i,Y_k) Y_i^\top Y_j Y_i^\top \rho_2(Y_j,Y_\ell)
= \textnormal{tr} \! \left(\! \Sigma_n \left( \frac{1}{n} \! \sum_j \overline{V}_{n,2}(Y_j) Y_j^\top \! \right) \! \right),
\end{eqnarray*}
\begin{eqnarray*}
\iint  \textnormal{I}_n(s) t^\top  \textnormal{C}_n(t)
  \textnormal{I}_n(s) \textnormal{I}_n(t) w(s,t)\, \textrm{{\rm d}}s\textrm{{\rm d}}t
  \! & \! = \! & \! \left(\! \frac{1}{n^2} \sum_{i,k} \rho_1(Y_i,Y_k)\! \right) \!
  \left(\! \frac{1}{n^2}\! \sum_{j,\ell} Y_j^\top \rho_2(Y_j,Y_\ell)\! \right)\\
 \! & \! = \! & \! V_{n,1} \, \frac{1}{n} \sum_j Y_j^\top \overline{V}_{n,2}(Y_j),
\end{eqnarray*}
\begin{eqnarray*}
& & \iint  s^\top \textnormal{R}_n(s) \frac{1}{n} \sum_j Y_j t^\top Y_j Y_j^\top \textnormal{C}_n(t)
  \textnormal{I}_n(s) \textnormal{I}_n(t) \,
w(s,t)\, \textrm{{\rm d}}s\textrm{{\rm d}}t \\
& = & \left( \frac{1}{n^2} \sum_{i,\ell} \rho_2(Y_i,Y_\ell)\right)^\top \left(\frac{1}{n^3}\sum_{j,k,m} Y_jY_j^\top Y_kY_j^\top \rho_2(Y_k,Y_m)\right)\\
& = & V_{n,2}^\top \left(\frac{1}{n^2}\sum_{j,k} Y_jY_j^\top Y_kY_j^\top \overline{V}_{n,2}(Y_k)\right),
\end{eqnarray*}
\[ \iint  \textnormal{C}_n(s)^\top  \frac{1}{n} \sum_j Y_j Y_j^\top s t^\top Y_jY_j^\top \textnormal{C}_n(t)
  \, \textnormal{I}_n(s) \textnormal{I}_n(t) \,
w(s,t)\, \textrm{{\rm d}}s\textrm{{\rm d}}t  =  \frac{1}{n} \sum_j \left(Y_j^\top \Gamma_n Y_j\right)^2,
\]
and
\[
\iint  s^\top \textnormal{C}_n(s) t^\top \textnormal{C}_n(t)  \, \textnormal{I}_n(s) \textnormal{I}_n(t) \,
w(s,t)\, \textrm{{\rm d}}s\textrm{{\rm d}}t
=  \left( \frac{1}{n^2} \sum_{i,k} Y_i^\top \rho_2(Y_i,Y_k)\right)^2.
\]
Here, summation is from 1 to $n$ for each of the indices, and each integral is over $\R^d$. \qed

\vspace*{3mm}

{\sc Proof of Theorem \ref{thmconsist}.}  The first observation is the following: Put
\begin{equation*}
\widehat{\sigma}_{n,0}^2 = 4 \int_{\R^d}\int_{\R^d}K^0_n(s,t) \, \textnormal{I}^0_n(s) \textnormal{I}^0_n(t) \,
w(s,t)\, \textrm{{\rm d}}s\textrm{{\rm d}}t,
\end{equation*}
where $K^0_n(s,t)$ originates from  $K(s,t)$ by replacing the functions $\textnormal{R}(\cdot)$, $\textnormal{I}(\cdot)$, $\textnormal{C}(\cdot)$ and $\textnormal{S}(\cdot)$ with their respective 'estimator-free' empirical counterparts
\begin{eqnarray*}
\textnormal{R}^0_n(t) & = & \frac{1}{n} \sum_{j=1}^n \cos\left(t^\top X_j\right), \quad
\textnormal{I}^0_n(t) = \frac{1}{n} \sum_{j=1}^n \sin\left(t^\top X_j\right),\\
\textnormal{C}^0_n(t) & = & \frac{1}{n} \sum_{j=1}^n X_j\cos\left(t^\top X_j\right), \quad \textnormal{S}^0_n(t)
 = \frac{1}{n} \sum_{j=1}^n X_j \sin\left(t^\top X_j\right), \ \ t \in \R^d,
\end{eqnarray*}
and do the same with each of the  five explicitly designated expectations figuring in the definition of $K(s,t)$.
Hence $\BE \left[\sin\left(s^\top X\right)\sin\left(t^\top X\right)\right]$ is replaced with $n^{-1}\sum_{j=1}^n \!\sin(s^\top \! X_j)$
$\sin(t^\top X_j)$ etc.
It is then straightforward to see that
\begin{equation}\label{convsigma0}
\widehat{\sigma}_{n,0}^2 \overset{\PP}{\longrightarrow} \sigma^2.
\end{equation}
For example, apart from the factor 4, the contribution of the first summand of the representation of $K(s,t)$ to $\sigma^2$ is
\[
J := \iint \BE\left[\sin(s^\top X)\sin(t^\top X)\right] \textnormal{I}(s)\textnormal{I}(t) \, w(s,t) \, \textnormal{d}s\textnormal{d}t
\]
(say). For the empirical version
\[
J_n = \iint \frac{1}{n}\sum_i \sin(s^\top X_i)\sin(t^\top X_i) \, \textnormal{I}^0_n(s) \textnormal{I}^0_n(t)
\, w(s,t) \, \textnormal{d}s\textnormal{d}t
\]
(say), Fubini's theorem gives
\[
\BE(J_n) = \frac{1}{n^3} \sum_{i,j,k} \iint \BE\left[ \sin(s^\top X_i)\sin(t^\top X_i)\sin(s^\top X_j)\sin(t^\top X_k)\right]
\, w(s,t) \, \textnormal{d}s\textnormal{d}t.
\]
If all indices are different, then, by symmetry and independence, the expectation beneath the integral sign is
$\BE[\sin(s^\top X)\sin(t^\top X)] \textnormal{I}(s)\textnormal{I}(t)$. Since the case that at least two of the three indices coincide is
asymptotically negligible, we have $\lim_{n\to \infty} \BE(J_n) = J$. Likewise, $\lim_{n\to \infty} \BV(J_n) =0$ and thus
 $J_n \overset{\PP}{\longrightarrow} J$. Since the other terms can be treated similarly, \eqref{convsigma0} follows.

 The much more difficult part of the proof is to show
 \begin{equation}\label{diffsigma}
 \widehat{\sigma}_n^2 - \widehat{\sigma}_{n,0}^2 \overset{\PP}{\longrightarrow} 0.
  \end{equation}
  In view of the definitions of $\widehat{\sigma}_n^2$ and $\widehat{\sigma}_{n,0}^2$, this boils down to prove
  \[
  \iint \big{(} K_n(s,t) \textnormal{I}_n(s) \textnormal{I}_n(s) -  K^0_n(s,t) \textnormal{I}^0_n(s) \textnormal{I}^0_n(s) \big{)} w(s,t)
  \, \textnormal{d}s\textnormal{d} t  \overset{\PP}{\longrightarrow} 0.
  \]
  To this end, we have to consider each term of the various summands comprising $K_n(s,t)$ and compare this with the corresponding term in
  $K^0_n(s,t)$. As an example, we choose the empirical versions of the first summand of $K(s,t)$ that involves  moments of $X$ which, apart form the minus sign and the factor $1/2$, is $t^\top \BE[\sin(s^\top X)XX^\top]$. Putting
  \begin{eqnarray*}
  L_n(s,t) \! & \! = \! & \! t^\top \frac{1}{n^4} \sum_{j,k,\ell,m} \sin(s^\top Y_j)Y_j Y_j^\top \, Y_k \cos(t^\top Y_k) \, \sin(s^\top Y_\ell)
\,  \sin(t^\top Y_m),\\
  L^0_n(s,t) \! & \! = \! & \! t^\top \frac{1}{n^4} \sum_{j,k,\ell,m}  \sin(s^\top X_j)X_j X_j^\top \,  X_k \cos(t^\top X_k) \, \sin(s^\top X_\ell)
  \,  \sin(t^\top X_m),\\
  \end{eqnarray*}
  we have to prove
  \begin{equation*}
  \iint \big{(} L_n(s,t) - L^0_n(s,t) \big{)} w(s,t) \, \textnormal{d}s \textnormal{d}t  \overset{\PP}{\longrightarrow} 0 .
  \end{equation*}
Notice that
\[
\iint L_n(s,t) \, w(s,t) \, \textnormal{d}s \textnormal{d}t = \frac{1}{n^4} \sum_{j,k,\ell,m} D(Y_j,Y_\ell) \, E(Y_j,Y_k,Y_m),
\]
where
\begin{eqnarray*}
D(Y_j,Y_\ell) & = & \int \sin(t^\top Y_j)\sin(s^\top Y_\ell) \, \exp(-a\|s\|^2) \, \textnormal{d}s,\\
E(Y_j,Y_k,Y_m) & = & \int t^\top Y_jY_j^\top Y_k \cos(t^\top Y_k) \sin(t^\top Y_m) \, \exp(-a\|t\|^2) \, \textnormal{d}t.
\end{eqnarray*}
Likewise,
\[
\iint L^0_n(s,t) \, w(s,t) \, \textnormal{d}s \textnormal{d}t = \frac{1}{n^4} \sum_{j,k,\ell,m} D(X_j,X_\ell) \, E(X_j,X_k,X_m),
\]
where
\begin{eqnarray*}
D(X_j,X_\ell) & = & \int \sin(t^\top X_j)\sin(s^\top X_\ell) \, \exp(-a\|s\|^2) \, \textnormal{d}s,\\
E(X_j,X_k,X_m) & = & \int t^\top X_jX_j^\top X_k \cos(t^\top X_k) \sin(t^\top X_m) \, \exp(-a\|t\|^2) \, \textnormal{d}t.
\end{eqnarray*}
It follows that
\[
\iint \big{(} L_n(s,t) - L^0_n(s,t) \big{)} w(s,t) \, \textnormal{d}s \textnormal{d}t = K_{n,1} + K_{n,2},
\]
where
\begin{eqnarray*}
K_{n,1} & = & \frac{1}{n^4} \sum_{j,k,\ell,m} \Big{(} E(Y_j,Y_k,Y_m) - E(X_j,X_k,X_m)\Big{)} D(Y_j,Y_\ell), \\ \label{zeile2}
K_{n,2} & = & \frac{1}{n^4} \sum_{j,k,\ell,m} E(X_j,X_k,X_m) \Big{(} D(Y_j,Y_\ell) - D(X_j,X_\ell)\Big{)}.
\end{eqnarray*}
We first prove $K_{n,2} \overset{\PP}{\longrightarrow} 0$.
Putting
\[
c_\nu := \int \|t\|^\nu \, \exp(-a\|t\|^2) \, \textrm{d}t, \quad \nu \in \{1,2\},
\]
the fact that $|\cos(t^\top X_k) \sin(t^\top X_m)| \le 1$ and the Cauchy-Schwarz inequality yield
\[
|E(X_j,X_k,X_m)| \le c_1 \, \|X_j\|^2 \, \|X_k\|.
\]
Since $\sin(t^\top Y_j) = \sin(t^\top X_j) + \xi_j t^\top \Delta_j$, where $|\xi_j| \le 1$ (and likewise for
$\sin(t^\top Y_\ell)$), the Cauchy-Schwarz inequality gives
\begin{eqnarray*}
|D(Y_j,Y_\ell) - D(X_j,X_\ell)| & \le & \int \big{(}\|t\| (\| \Delta_j\| + \|\Delta_\ell\|) + \|t\|^2 \| \Delta_j \| \|\Delta_\ell \|\big{)}e^{-a\|t\|^2} \textnormal{d}t\\
& = & c_1\left(\|\Delta_j\| + \|\Delta_\ell\|\right) + c_2 \|\Delta_j\| \|\Delta_\ell\|.
\end{eqnarray*}
We therefore have
\[
|K_{n,2}| \le \frac{1}{n^4} \sum_{j,k,\ell,m} c_1 \|X_j\|^2 \|X_k\| \Big{(} c_1(\|\Delta_j\| + \|\Delta_\ell\|) + c_2 \|\Delta_j\| \|\Delta_\ell\| \Big{)}.
\]
Since $n^{-1}\sum_{j=1}^n \|X_j\|^\nu = O_\PP(1)$ if $\nu \in \{1,2,3,4\}$ (recall the assumption $\BE \|X\|^4 < \infty$) and
\[
\frac{1}{n} \sum_{j=1}^n \|X_j\|^2 \|\Delta_j\| \le \left( \frac{1}{n} \sum_{j=1}^n \|X_j\|^4 \cdot \frac{1}{n} \sum_{j=1}^n \|\Delta_j\|^2\right)^{1/2},
\]
$K_{n,2} \overset{\PP}{\longrightarrow} 0$ follows from \eqref{deltabound} and \eqref{deltanquer}.

As for $K_{n,1}$, first notice that $|D(Y_j,Y_\ell)| \le c_1$ and thus
\[
|K_{n,1}| \le \frac{c_1}{n^4} \sum_{j,k,\ell,m} | E(Y_j,Y_k,Y_m) - E(X_j,X_k,X_m)|.
\]
 Next, we have
\[
Y_j Y_j^\top Y_k  =  X_j X_j^\top X_k + \widetilde{\Delta}_{j,k},
\]
where
\[
\widetilde{\Delta}_{j,k}  =  X_j\Delta_j^\top X_k + X_jX_j^\top\Delta_k + X_j\Delta_j^\top \Delta_k + \Delta_j X_j^\top X_k
 + \Delta_j \Delta_j^\top X_k + \Delta_j X_j^\top \Delta_k + \Delta_j\Delta_j^\top \Delta_k.
\]
Therefore,
\begin{eqnarray}\nonumber
\! \! & \! \! \! \!  & \! \!  E(Y_j,Y_k,Y_m) - E(X_j,X_k,X_m) \\ \label{line1}
\! \! & \! \! = \! \! & \! \! \int \! t^\top X_jX_j^\top X_k \Big{(} \! \cos(t^\top \! Y_k) \sin(t^\top \! Y_m) - \cos(t^\top \! X_k) \sin(t^\top \! X_m) \! \Big{)} e^{-a\|t\|^2} \textnormal{d} t\\ \nonumber
\! \! & \! \! \! \! & \! \! + \int t^\top \widetilde{\Delta}_{j,k} \cos(t^\top Y_k) \sin(t^\top Y_m) e^{-a\|t\|^2} \textnormal{d} t.
\end{eqnarray}
Since
\begin{eqnarray*}
\|\widetilde{\Delta}_{j,k}\| & \le & \|X_j\| \| \Delta_j\| \|X_k\|  + \|X_j\|^2 \|\Delta_k\| + \|X_j\| \|\Delta_j\| \Delta_k\|
 + \| \Delta_j\| \|X_j\| \|X_k\|\\
 & &  + \|\Delta_j\|^2 \|X_k\| +  \|\Delta_j\| \|X_j\| \|\Delta_k\| + \|\Delta_j\|^2 \|\Delta_k\|,
 \end{eqnarray*}
 the inequality $|\cos(t^\top Y_k) \sin(t^\top Y_m)| \le 1$ and the same reasoning as above show that
 \[
 \frac{1}{n^4} \sum_{j,k,\ell,m} \bigg{|} \int t^\top \widetilde{\Delta}_{j,k} \cos(t^\top Y_k) \sin(t^\top Y_m) e^{-a\|t\|^2} \textnormal{d} t\bigg{|} = o_\PP(1).
 \]
  Regarding the term figuring in \eqref{line1}, we have
 \[
 |\cos(t^\top \! Y_k) \sin(t^\top \! Y_m) - \cos(t^\top \! X_k) \sin(t^\top \! X_m)| \le \|t\|(\|\Delta_k\| + \|\Delta_m\|) + \|t\|^2 \|\Delta_k\|\|\Delta_m\|,
 \]
 and it  follows by the same reasoning as above that
 \[
 \frac{1}{n^4} \sum_{j,k,\ell,m} \bigg{|} \int \! t^\top X_jX_j^\top X_k \Big{(} \! \cos(t^\top \! Y_k) \sin(t^\top \! Y_m) - \cos(t^\top \! X_k) \sin(t^\top \! X_m) \! \Big{)} e^{-a\|t\|^2} \textnormal{d} t \bigg{|} = o_\PP(1).
  \]
 Consequently, $K_{n,1} = o_\PP(1)$. Since all the other summands comprising $K_n$ and $K^0_n$ can be tackled in the same way,
 \eqref{diffsigma} follows. \qed

\end{document}